\documentclass[%
twocolumn,
physplasmas,
aip,
 amsmath,amssymb,
reprint,%
]{revtex4-2}

\usepackage{graphicx}
\usepackage{dcolumn}
\usepackage{bm}
\usepackage{color}

\begin{document}

\preprint{AIP/123-QED}


\title{
%
%
Linear 
Landau damping, Schr\"{o}dinger equation, and fluctuation theorem
}

\author{H. Sugama}

\affiliation{
National Institute for Fusion Science, 
Toki 509-5292, Japan
}
\affiliation{
Department of Advanced Energy, University of Tokyo, 
Kashiwa 277-8561, Japan
}

\date{\today}

\begin{abstract}
A linearized Vlasov-Poisson system of equations is transformed into a Schr\"{o}dinger equation, which is used to demonstrate that the fluctuation theorem holds for the relative stochastic entropy, defined in terms of the probability density functional of the particle velocity distribution function in the Landau damping process. 
The difference between the energy perturbation, normalized by the equilibrium temperature, and the entropy perturbation constitutes a time-independent invariant of the system. 
This invariant takes the quadratic form of the perturbed velocity distribution function and corresponds to the squared amplitude of the state vector that satisfies the Schr\"{o}dinger equation. 
%
%
Exact solutions, constructed from a discrete set of Hamiltonian eigenvectors, are employed to formulate and numerically validate the fluctuation theorem for the Landau damping process. 
The results offer new insights into the formulations of collisionless plasma processes within the framework of nonequilibrium statistical mechanics.

\end{abstract}

\maketitle

Landau damping~\cite{Landau} has been extensively studied as one of the primary physical mechanisms responsible for stabilizing microinstabilities and zonal-flow oscillations, as well as for wave heating in high-temperature plasmas, such as those found in space and fusion devices~\cite{Case,VK,Nicholson,H&P,Sugama1999,Sugama2006,Biancalani,Loureiro,Plunk,Maekaku}. 
It is a seemingly irreversible process, despite occurring in a collisionless plasma governed by the Vlasov equation with time-reversal symmetry.
On the other hand, the fluctuation theorem~\cite{FTES,Jarzynski,Shiraishi}, derived from reversible dynamics, states that the probability ratio of entropy production to reduction grows exponentially with time, thus providing a microscopic foundation for the second law of thermodynamics and nonequilibrium statistical mechanics.
%
This Letter reformulates a linearized Vlasov-Poisson system of equations as a Schr\"{o}dinger equation to concisely capture the properties of conservation and time reversibility, thereby enabling the application of the fluctuation theorem to the Landau damping process.
The same Hermite expansion form of a Schr\"{o}dinger equation as in the present work 
was derived by Ameri, {\it et al}.~\cite{Ameri}
who presented the quantum algorithm for solving the linear Vlasov-Poisson system. 
In contrast to their work, this study provides  
a representation of the Schr\"{o}dinger equation in terms of the eigenvectors 
corresponding to the Case-Van Kampen modes~\cite{Case,VK,Nicholson}. 
This novel framework offers several key advancements, including the definition of stochastic relative entropy, the derivation of the fluctuation theorem, and its numerical verification, thereby providing deeper insights into Landau damping from the perspective of nonequilibrium statistical mechanics.
%


%
The distribution function of electrons in a two-dimensional phase space at time $t$ is denoted 
by $f(x, v, t)$,  
and 
$f(x, v, t) dx dv$ represents
the number of electrons whose positions and velocities lie within 
the infinitesimal intervals 
$[x, x + dx)$ and $[v, v+ dv)$, respectively. 
%
%
%
In a collisionless system, $f(x, v, t)$ is governed by 
the Vlasov equation~\cite{Nicholson},
\begin{equation}
\frac{\partial f(x, v, t)}{\partial t}
+
v
\frac{\partial f(x, v, t)}{\partial x}
- 
\frac{e}{m} E(x, t) 
\frac{\partial f(x, v, t)}{\partial v}
=
0
, 
\end{equation}
where $m$ and $-e$ are the electron mass and charge, respectively. 
%
%
The motion of ions, which are assumed to have a uniform density $n_0$, is neglected 
on the ground that the ion mass is much larger than the electron mass.
%
The electric field $E(x, t)$ in the $x$-direction is determined from $f(x, v, t)$
through Poisson's equation, 
$
\partial E / \partial x
= 4 \pi e
(
n_0 
- \int_{-\infty}^{+\infty} d v f
)
$. 
The system is assumed to be periodic with period length $L$ in the $x$-direction,
and the constraint condition,  $\int_{-L/2}^{L/2} dx E(x,t) = 0$,  is imposed. 
%
Here, we do not consider an equilibrium electric field that could 
give rise to an inhomogeneous equilibrium distribution of electrons.
%
The nonlinear Vlasov-Poisson system described above conserves 
the energy 
$
{\cal E}
\equiv
(n_0 L)^{-1}
\int_{-L/2}^{L/2} dx \,
[
\int_{-\infty}^{+\infty} dv 
f m v^2/ 2
+
E^2/ 8\pi 
]
$
and the Gibbs entropy
$
S_f
\equiv
-
(n_0 L)^{-1}
\int_{-L/2}^{L/2} dx \int_{-\infty}^{+\infty} dv 
f \log f
$,
both defined per single electron.

The distribution function $f(x, v, t)$ is assumed to consist of a Maxwellian equilibrium part 
$
f_0 (v)
= 
\pi^{-1/2} (n_0/v_T)
\exp
( - v^2/v_T^2 )
$ 
and 
a perturbed part $f_1(x, v, t)$. 
Here, 
$
v_T \equiv \sqrt{2} v_t \equiv 
\sqrt{2T/m}
$ 
where $T$ is the electron temperature. 
Hereafter, we neglect the nonlinear term 
$-  (e / m) E(x, t) \partial f_1(x, v, t) / \partial v$ in the Vlasov equation. 
%
The assumption of linearity for the Vlasov-Poisson system is essential for deriving the Schr\"{o}dinger with the spectral presentation of the state vector and the Hamiltonian operator as shown later.
%
It can be shown that 
\begin{equation}
D[f_1]
\equiv 
\int_{-L/2}^{+L/2}\frac{dx}{L}
\biggl[
 \frac{ [E(x, t)]^2 }{8 \pi n_0 T}
+ 
\frac{1}{n_0}
\int_{-\infty}^{+\infty} dv \; \frac{ [f_1(x,v,t)]^2 }{2 f_0(v)}
\biggr]
, 
\end{equation}
is rigorously time-independent for any solution $f_1$ of the linearized Vlasov-Poisson equations. 
The invariant functional $D[f_1]$ takes a quadratic form with respect to $f_1$ 
and satisfies the relation,  
$
D[f_1] =   {\cal E}^{(2)} / T - S_f^{(2)} 
$, 
where ${\cal E}^{(2)}$ and $S_f^{(2)}$ represent the second-order terms in the expansions of ${\cal E}$ and  $S_f$, 
respectively, 
with respect to the ordering parameter $\alpha \sim f_1/f_0$, 
which characterizes the perturbation amplitude~\cite{Maekaku}. 
%
We note that 
the neglected nonlinear term 
drives the long-time-scale evolution of the equilibrium (or background) distribution function as 
a quasilinear effect of ${\cal O}(\alpha^2)$ which can be evaluated 
using the ${\cal O}(\alpha)$ solutions for $E(x,t)$ and $f_1(x,v,t)$ obtained from the linear theory 
as shown in Ref.~\cite{Maekaku}. 
Also, note that,  
although 
${\cal E}^{(2)}$ and  $S_f^{(2)}$ are not separately conserved in the linear system, 
it inherits a certain conservation property
from the original nonlinear system, as indicated by the conservation of  $D[f_1]$. 
%


We now assume 
$ f_1 (x, v, t)$ to be expressed as 
$
f_1 (x, v, t) = \mbox{Re}[ f_1(k, v, t) \exp ( i k  x ) ]
$
with the wavenumber $k= 2\pi/ L > 0$. 
%
Here, 
only the mode with $k=2\pi/L$ is considered, and higher-order harmonics are not included. 
%
The normalized time and velocity are defined by 
$\tau \equiv k v_T t$
and $\xi \equiv  v / v_T$, respectively.
Using the Hermite polynomials
$
H_n(\xi) \equiv (-1)^n e^{\xi^2} d^n (e^{-\xi^2}) /d\xi^n 
\;
( n = 0, 1,  2, \cdots )
$, 
we here define the functions 
$
h_n (\xi)
\equiv
\pi^{-1/4} e^{-\xi^2/2}
 H_n(\xi) /  
 (2^n n!)^{1/2}
$
which satisfy the orthonormality condition, 
$
\int_{-\infty}^{+\infty} d\xi \;  
h_n (\xi) h_{n'} (\xi) 
= \delta_{n n'}
$.
Then,  $f_1(k, v, t)$ is 
represented by the dimensionless function $\widetilde{f} (\kappa, \xi, \tau)$ as 
$
f_1(k, v, t) 
= (n_0/v_T) h_0 (\xi ) \widetilde{f} (\kappa, \xi, \tau) 
$
where the normalized wavenumber is defined by $\kappa \equiv k \lambda_D$ 
with the Debye length $\lambda_D \equiv \omega_p/ v_t$ 
and the plasma frequency $\omega_p = (4\pi n_0 e^2/m)^{1/2}$. 

We associate complex-valued functions of the normalized velocity variable $\xi$ 
with ket vectors denoted by the symbol $| \;  \rangle$, following the notation in quantum mechanics~\cite{QM}. 
%
Here, 
the dependence of the perturbed distribution function on the position variable $x$ is 
specified as $\propto \exp ( i k x)$, so we focus on the space of functions that depend only on 
the velocity variable $\xi = v / v_T$. 
Thus, a ket vector describes the electron distribution in velocity space 
in contrast to a standard quantum mechanical wave function, which represents 
the particle distribution in position space. 
%
The ket vectors $| n \rangle$ and $|\xi'\rangle$ correspond to   
the function $h_n (\xi)$ and the delta function $\delta (\xi -\xi')$, 
respectively. 
%
%
The bra vector conjugate to $|u\rangle$ is denoted by $\langle u |$. 
 Then,  $\delta (v-v')$ and $h_n (\xi)$ are expressed through the scalar products as 
$
\langle \xi | \xi' \rangle 
= 
\delta (\xi - \xi')
$
and 
$
\langle \xi | n \rangle 
= 
h_n (\xi) 
$, 
and 
the orthonormality condition satisfied by $h_n (\xi)$ is written as 
$
\langle n | n' \rangle 
= 
\int_{-\infty}^{+\infty} d\xi \;  
\langle n | \xi \rangle \langle \xi | n' \rangle 
= \delta_{n n'}
$. 
Note that $\{ | \xi  \rangle \}_{-\infty < \xi < + \infty}$ and 
$\{ | n  \rangle \}_{n = 0, 1, 2, \cdots}$ constitute two distinct sets of 
orthonormal basis vectors that satisfy the closure relation, 
$
\int_{-\infty}^{+\infty} 
| \xi \rangle d\xi \langle \xi | 
= 
\sum_{n=0}^\infty
| n \rangle  \langle n | 
= 
\widehat{1}
$,
where $\widehat{1}$ is the identity operator. 
%
%
The operators $\widehat{\Xi}$ and $\widehat{N}$ are defined by 
$
\widehat{\Xi}
\equiv 
\int_{-\infty}^{+\infty} d\xi \; | \xi \rangle 
\xi d \xi \langle \xi |
$,
and 
$
\widehat{N}
\equiv 
\sum_{n=0}^\infty  | n \rangle 
n \langle n |
$,
from which it follows that
$
\widehat{\Xi} | \xi \rangle 
= 
\xi | \xi \rangle 
$
and 
$
\widehat{N}  | n \rangle 
= 
n  | n \rangle 
$.
%
A representation of state vectors and operators refers to expressing them as column vectors and matrices of complex numbers with respect to a chosen set of basis vectors, and it depends on that choice.~\cite{QM}
In quantum mechanics, it is common to use the eigenvectors of a certain Hermitian operator as orthonormal basis vectors for a representation. 
Two representations associated with 
the aforementioned sets of basis vectors $\{ | \xi  \rangle \}_{-\infty < \xi < + \infty}$ and $\{ | n  \rangle \}_{n = 0, 1, 2, \cdots}$  are referred to as the $\{ \Xi \}$ and $\{ N \}$ representations. 
%

We now consider the ket vector $|\widetilde{f} (\tau) \rangle$, which 
is a function of $\tau$ and related to 
the perturbed distribution function $\widetilde{f} (\kappa, \xi, \tau) $ 
by 
$
\langle \xi  | \widetilde{f} (\tau) \rangle
= 
\widetilde{f} (\kappa, \xi, \tau) 
$, 
where the $\kappa$-dependence is omitted in the notation $| \widetilde{f} (\tau) \rangle$ for simplicity.  
We also define the Hermitian operator $\widehat{A}$ by
\begin{equation}
\widehat{A}
=
\widehat{1}
+ 
| 0 \rangle
\bigl[
\left( 1 + \kappa^{-2} \right)^{1/2} - 1
\bigr]
\langle 0 | 
. 
\end{equation}
Defining the state vector 
$
| \psi (\tau) \rangle \equiv 
\widehat{A} |\widetilde{f}(\tau)\rangle
$,
the invariant $D[f_1]$ can be expressed as 
\begin{equation}
D[f_1]
= 
 \frac{1}{4}
\langle \widetilde{f} (\tau) | 
\widehat{A}^2
 | \widetilde{f} (\tau) \rangle
 = 
  \frac{1}{4}
\langle \psi (\tau) 
 | \psi (\tau) \rangle
.
\end{equation}
It follows that 
$\langle \psi (\tau) | \psi (\tau) \rangle$ is independent of $\tau$, 
and 
the time evolution operator $\widehat{U}(\tau)$ defined by 
$
 | \psi (\tau) \rangle
= 
\widehat{U}(\tau)
 | \psi (0) \rangle
$
is unitary. 
Here, $\widehat{U}(\tau)$ can be written as 
$
\widehat{U} (\tau)
= 
\exp ( - i \tau \widehat{H} )
$, 
where $\widehat{H}$ is the Hamiltonian operator defined by 
$
\widehat{H} 
=
\widehat{A} \, \widehat{\Xi} \, \widehat{A}
$. 
Thus, 
$\widehat{H}$ is Hermitian, 
and $| \psi (\tau) \rangle$ satisfies 
\begin{equation}
i \frac{d}{d \tau}
| \psi (\tau) \rangle
= 
\widehat{H} 
| \psi (\tau) \rangle
, 
\end{equation}
which takes the form of the Schr\"{o}dinger equation 
with $\hbar = 1$.
%
As shown above,
the time evolution of the perturbed distribution function $f_1(k, v, t)$ is mapped to
that of the state vector $|\psi(\tau)\rangle$.
Accordingly, the Schr\"{o}dinger picture of quantum mechanics is employed here, rather than
the Heisenberg picture.
This quantum mechanical framework, which naturally incorporates the conservation law and time-reversal symmetry,
facilitates the formulation of the fluctuation theorem for the Landau damping process,
as demonstrated below.
%

%
%
In the $\{ N \}$ representation, 
the Hamiltonian $\widehat{H}$ and the Schr\"{o}dinger equation are written as 
\begin{equation}
\label{HN}
\widehat{H}
=
\frac{1}{\sqrt{2}}
\sum_{n=0}^\infty 
\sqrt{n + 1 + \kappa^{-2} \delta_{n0} }
\Bigl(
| n + 1 \rangle \langle n | + | n \rangle \langle n + 1 |
\Bigr)
, 
\end{equation}
and
$
i  \, 
d \psi_n (\tau)  / d\tau
=
\sum_{n'=0}^\infty H_{n n'} \psi_{n'} (\tau) 
$,
respectively, 
where 
$ \psi_n (\tau)  \equiv \langle n | \psi (\tau) \rangle$ 
and
$
 H_{n n'} 
 \equiv
\langle n | \widehat{H} | n' \rangle
$. 
%
The Hamiltonian $\widehat{H}$ in Eq.~(\ref{HN}) includes 
$| n+1 \rangle \langle n |$ and  $| n \rangle \langle n+1 |$ 
which play the roles of the creation and annihilation operators, respectively, for the 
basis vectors $|n\rangle$ $(n=0, 1, 2, \cdots)$. 
Therefore,  although the present system is intrinsically classical, 
Landau damping can be interpreted as an energy transfer process 
from macroscopic to microscopic structures in the velocity-space 
distribution function, mediated by the creation and annihilation 
of the discrete states $|n\rangle$ $(n=0, 1, 2, \cdots)$, which resemble those 
of the quantum harmonic oscillator. 
%


%
The eigenvectors of the Hamiltonian $\widehat{H}$ 
are derived from the Case-Van Kampen modes~\cite{Case,VK,Nicholson} as shown below. 
%
The perturbed distribution function for 
the Case-Van Kampen mode is given by 
$f_{{\rm CVK}, \zeta} (k, v, t) \equiv (n_0/v_T) h_0(\xi) 
\widetilde{f}_{{\rm CVK}, \zeta}(\kappa, \xi, \tau)$ 
for 
$-\infty < \zeta < + \infty$, 
where $\widetilde{f}_{{\rm CVK}, \zeta}(\kappa, \xi, \tau)$ 
is expressed in terms of $| \widetilde{f}_{{\rm CVK}, \zeta} \rangle$ 
as 
\begin{eqnarray}
 & & 
  \widetilde{f}_{{\rm CVK}, \zeta}(\kappa, \xi, \tau)
   \equiv  
 \langle \xi | \widetilde{f}_{{\rm CVK}, \zeta} \rangle
\nonumber 
\\ & & 
\hspace*{-3mm}
= \frac{1}{h_0 (\xi)} 
\left[ 
\delta ( \xi  - \zeta )  \, {\rm Re} [ \epsilon (\xi)]
-  \frac{1}{\pi} 
P \left( \frac{1}{\xi - \zeta} \right) 
\, {\rm Im} [ \epsilon (\xi)] 
\right]
. 
\hspace*{2mm}
\end{eqnarray}
Here, 
$
\epsilon(\zeta)
 \equiv 
1 + \kappa^{-2}
[ 1 + \zeta Z(\zeta) ]
$
and $Z(\zeta)$ is the plasma dispersion function that is defined by 
$
Z(\zeta)
=  \pi^{-1/2} P
\int_{-\infty}^{+\infty} dz \,
e^{-z^2}/ (z - \zeta)
+ i \pi^{1/2} e^{-\zeta^2}
$
for a real number $\zeta$. 
It is found that the CVK state vector defined by 
\begin{equation}
| {\rm CVK}, \zeta \rangle
\equiv 
\frac{h_0(\zeta)} {|\epsilon(\zeta)|}
\widehat{A} 
| \widetilde{f}_{{\rm CVK}, \zeta} \rangle
\end{equation}
%
satisfies the eigenvector equation for $\widehat{H}$ with the eigenvalue $\zeta$: 
%
$
\widehat{H} 
| {\rm CVK}, \zeta \rangle
\equiv
\zeta 
| {\rm CVK}, \zeta \rangle
$. 
Then, 
$| {\rm CVK}, \zeta \rangle$ $(-\infty < \zeta < + \infty)$
form a complete orthonormal basis, satisfying 
$
\langle {\rm CVK}, \zeta
| {\rm CVK}, \zeta' \rangle
= 
 \delta (\zeta - \zeta') 
$ 
and 
$
\int_{-\infty}^{+\infty}
| {\rm CVK}, \zeta \rangle
d\zeta
\langle {\rm CVK}, \zeta |
= 
\widehat{1}
$. 
In the $\{ \rm{CVK} \}$ representation associated 
with the CVK basis vectors, 
the Hamiltonian 
and 
the time evolution operator are diagonalized as 
$
\widehat{H}
= 
\int_{-\infty}^{+\infty}
| {\rm CVK}, \zeta \rangle
\zeta  d\zeta
\langle {\rm CVK}, \zeta |
$
and 
$
 \widehat{U} (\tau)
= 
\int_{-\infty}^{+\infty}
| {\rm CVK}, \zeta \rangle
e^{-i \zeta \tau} d\zeta
\langle {\rm CVK}, \zeta |
$, 
respectively. 
%
%
%
%
%


We consider a finite set of 
CVK state vectors,  
$\{ | {\rm CVK}, \zeta_j \rangle \}_{j =0, 1, \cdots, N_{\rm cvk}-1 }$ 
for a given positive integer $N_{\rm cvk}$. 
Here, 
$\{ \zeta_j  \}_{j =0, 1, \cdots, N_{\rm cvk}-1 }$
represent $N_{\rm cvk}$ 
real-valued solutions of the $N_{\rm cvk}$th-order algebraic equation for $\zeta$ 
given by the condition, 
$
\langle N_{\rm cvk} | {\rm CVK}, \zeta \rangle = 0
$, 
where $\langle N_{\rm cvk} |$ is the $N_{\rm cvk}$th basis bra vector in the $\{ N \}$ representation. 
Then, instead of treating the full space of state vectors, 
we focus on the subspace spanned by the 
$N_{\rm cvk}$ CVK state vectors. 
This subspace is invariant under the action of the Hamiltonian $\widehat{H}$ and the time evolution 
operator $\widehat{U}(\tau)$ because the CVK state vectors are the eigenvectors of $\widehat{H}$. 
Any state vector in the subspace at time $\tau$ can be expressed as 
\begin{equation}
| \psi (\tau) \rangle
= \sum_{j=0}^{ N_{\rm cvk} - 1 }
c_j (\tau)  | {\rm CVK}, \zeta_j \rangle
, 
\end{equation}
where 
$
c_j (\tau) = c_j (0) \exp ( - i \zeta_j \tau )
$.
Thus, the components of $| \psi  (\tau) \rangle$ in the $\{ N \}$ representation 
are given by 
$
\psi_n (\tau) 
  \equiv  
\langle n 
| \psi (\tau) \rangle
=
 \sum_{j=0}^{ N_{\rm cvk} - 1 }
c_j (\tau)  \langle n | {\rm CVK}, \zeta_j \rangle
$. 
Note that 
$
\psi_{N_{\rm cvk}} (\tau) = 0
$
holds for any $\tau$. 
From the infinite set of components,  
we extract the first $N_{\rm cvk}$ components, 
$\{ \psi_n (\tau) \}_{n = 0, 1, \cdots, N_{\rm cvk} - 1}$, 
which form an $N_{\rm cvk}$-dimensional complex column vector, 
$
\boldsymbol{\psi} (\tau)
\equiv   
\mbox{}^t
[
\psi_0 (\tau), \psi_1 (\tau), \cdots, 
\psi_{N_{\rm cvk}-1} (\tau)
]
$, 
where $\mbox{}^t [\cdots]$ denotes the transpose of a row vector to express it as a column vector. 
There exists a one-to-one correspondence between such $N_{\rm cvk}$-dimensional complex vectors 
and the vectors in the subspace spanned by the $N_{\rm cvk}$ CVK state vectors. 
The vector 
$\boldsymbol{\psi} (\tau)$ 
satisfies 
the Schr\"{o}dinger equation, 
\begin{equation}
\label{SchN}
i \frac{d}{d \tau} \boldsymbol{\psi} (\tau)
=
{\bf H} \boldsymbol{\psi} (\tau)
,
\end{equation}
where ${\bf H} = [  H_{n n'}  ]_{n,n'= 0, 1, \cdots, N_{\rm cvk}-1}$ 
is a Hermitian $N_{\rm cvk} \times N_{\rm cvk}$ Hamiltonian matrix 
obtained as the submatrix of the infinite-dimensional matrix 
$[  H_{n n'}  ]_{n,n'= 0, 1, 2, \cdots}$ 
with the components defined by  
$
 H_{n n'} 
 \equiv
\langle n | \widehat{H} | n' \rangle
$ using $\widehat{H}$ in Eq.~(\ref{HN}). 
We can interpret $\boldsymbol{\psi} (\tau)$ as an approximate solution 
obtained by truncating the infinite-dimensional Schr\"{o}dinger equation 
in the $\{ N \}$ representation 
to a finite-dimensional system of size $N_{\rm cvk}$. 
At the same time, $\boldsymbol{\psi} (\tau)$ has a one-to-one correspondence to (and hence represents) 
the exact solution of the Schr\"{o}dinger equation in the state vector subspace spanned by the 
$N_{\rm cvk}$ CVK state vectors. 
The solution of Eq.~(\ref{SchN}) is given by 
$
\boldsymbol{\psi} (\tau)
=
{\bf U} (\tau) 
\boldsymbol{\psi} (0)
$
with the unitary matrix ${\bf U} (\tau) = \exp (-i \tau \mathbf{H})$.  
%
Thus, the squared norm, 
$
|| \boldsymbol{\psi} (\tau) ||^2
\equiv
\sum_{n=0}^{N_{\rm cvk} - 1} | \psi_n (\tau) |^2
$, 
remains constant in time $\tau$. 
%

We now assume the initial vector $\boldsymbol{\psi} (0)$
to be given randomly.  
Then, the vector $\boldsymbol{\psi} (\tau)$ at time $\tau$, that is uniquely 
determined from $\boldsymbol{\psi} (0)$, also becomes a random variable. 
Hereafter,  
$
\boldsymbol{\Psi} (\tau) \equiv   
\mbox{}^t
[
\Psi_0 (\tau), \Psi_1 (\tau), \cdots, 
\Psi_{N_{\rm cvk}-1} (\tau)
]
$
denotes the state vector as a random (or stochastic) variable while 
$
\boldsymbol{\psi} (\tau)
$ 
represents a specific realization of $\boldsymbol{\Psi} (\tau)$. 
%
More specifically, according to probability theory, the random variable 
$\boldsymbol{\Psi}(\tau)$ can also be regarded as a function of a hidden variable $\omega$ which represents the outcome of a random trial, and can be written as $\boldsymbol{\Psi}(\tau, \omega)$.
When $\omega$ takes a specific value as a result of the trial,
the realization of the random variable is expressed as
$\boldsymbol{\psi}(\tau) = \boldsymbol{\Psi}(\tau, \omega)$,
which represents the relation 
between the random variable $\boldsymbol{\Psi}$ and its realization $\boldsymbol{\psi}$.
%
%
The probability that the real and imaginary parts of the random variables
$\Psi_n (\tau)\equiv 
\Psi_{r,n} (\tau) + i \Psi_{i,n} (\tau)$ $(n=0, 1, 2, \cdots, N_{\rm cvk} - 1)$ lie
within 
the infinitesimal intervals 
%
$[\psi_{r,n}, \psi_{r,n} + d\psi_{r,n} )$ and 
$[\psi_{i,n}, \psi_{i,n} + d\psi_{i,n} )$, respectively, 
is given by 
$
P ( \boldsymbol{\psi} ; \tau )
\,
d \Gamma
$,
where 
the volume element is defined as  
$
d \Gamma 
\equiv
\prod_{n = 0}^{N_{\rm cvk}-1}
d\psi_{r,n} d\psi_{i,n}
$.
%
Since ${\bf U}(\tau)$ is the unitary matrix, 
$d \Gamma (\tau) \equiv
\prod_{n = 0}^{N_{\rm cvk}-1}
d\psi_{r,n} (\tau) d\psi_{i,n} (\tau)$
remains constant along the 
trajectory of the vector $\boldsymbol{\psi}(\tau)$ 
and therefore the probability density 
$
P [ \boldsymbol{\psi}(\tau) ; \tau] 
$
%
is independent of $\tau$. 
%
%
This corresponds to Liouville's theorem 
in Hamiltonian mechanics. 
%
Here, $P [ \boldsymbol{\psi}(\tau) ; \tau]$ 
denotes the value of the above-mentioned probability density $P ( \boldsymbol{\psi} ; \tau )$
evaluated at 
 $\boldsymbol{\psi} = \boldsymbol{\psi}(\tau)$ 
and time $\tau$. 
%

The stochastic relative entropy 
of the distribution 
$
P[ \boldsymbol{\Psi}(\tau) ; \tau ]
= 
P[ \boldsymbol{\Psi}(0) ; 0 ]
$ 
with respect to $P[ \boldsymbol{\Psi}(\tau) ; 0 ]$,  
%
which represents the initial probability density at the point $\boldsymbol{\Psi}(\tau)$ in the space of state vectors, 
%
is defined by 
\begin{equation}
\Delta S [ \boldsymbol{\Psi}(0) ; \tau ]
\equiv 
\log \left[
 \frac{  P [\boldsymbol{\Psi}(\tau); \tau ] }{
   P [ \boldsymbol{\Psi}(\tau) ; 0 ]
 }
\right]
\equiv
\log \left[
 \frac{  P [\boldsymbol{\Psi}(0); 0 ] }{
   P [ \boldsymbol{\Psi}(\tau) ; 0 ]
 }
\right]
,
\end{equation}
where $\boldsymbol{\Psi}(\tau)$ are related to 
$\boldsymbol{\Psi}(0)$
by 
$\boldsymbol{\Psi}(\tau) =  {\bf U}(\tau) \boldsymbol{\Psi}(0)$. 
%
Note that 
the difference between $P[\boldsymbol{\Psi}(\tau); \tau] = P[\boldsymbol{\Psi}(0);0]$ and 
$P[\boldsymbol{\Psi}(\tau); 0]$ causes $\Delta S [\boldsymbol{\Psi}(0) ; \tau]$ to become nonzero. 
%
We then define $P(\Delta S)$ as the probability density such that 
$P(\Delta S) d(\Delta S)$ gives 
the probability for the stochastic relative entropy 
$\Delta S [ \boldsymbol{\Psi}(0) ; \tau ] $  to 
take a value in the infinitesimal interval $[\Delta S, \Delta S + d(\Delta S) )$. 
The probability density $P(\Delta S)$ is given by 
\begin{equation}
P(\Delta S)
=
\int 
d \Gamma (0) \, 
%
 P[ \boldsymbol{\psi}(0) ; 0 ] 
%
   \delta [ \Delta S [ \boldsymbol{\psi}(0) ; \tau ] -  \Delta S  ]
.
\end{equation}
Now, assume that the initial probability density  
$P [ \boldsymbol{\psi}(0) ; 0 ]$ satisfies a symmetry condition, 
$
P [ {\bf T} \boldsymbol{\psi}(0); 0 ] 
= 
P [ \boldsymbol{\psi}(0) ; 0 ] 
$, 
%
where ${\bf T}$ is the diagonal matrix representing the time reversal transformation, 
defined by 
${\bf T} \equiv [ (-1)^i \delta_{ij}]_{i, j = 0, 1, 2, \cdots, N_{\rm cvk} - 1}$, 
which transforms the vector $\boldsymbol{\psi} = 
\mbox{}^t [\psi_0, \psi_1, \psi_2, \cdots, \psi_{N_{\rm cvk}-1} ]$ into 
${\bf T} \boldsymbol{\psi} = \mbox{}^t [\psi_0, - \psi_1, \psi_2, \cdots, (-1)^{N_{\rm cvk}-1} \psi_{N_{\rm cvk}-1} ]$. 
Noting that 
the perturbed distribution function is related to $\boldsymbol{\psi}$ by 
$f_1(k, v, t) 
= 
(n_0/v_T) \pi^{-1/2} e^{-\xi^2}
\sum_n (1+ \kappa^{-2} \delta_{n0} )^{-1/2} \psi_n (\tau) H_n (\xi) / (2^n n!)
$, 
we see that the transformation from the vector $\boldsymbol{\psi}(\tau)$ to ${\bf T} \boldsymbol{\psi}(\tau)$ 
corresponds to the transformation from the perturbed distribution function $f_1(k, v, t)$ to 
$f_1(k, - v, t)$. 
%
Then, following a procedure similar to that in Ref.~\cite{Jarzynski}, 
we can prove the fluctuation theorem, 
\begin{equation}
\label{FT}
\frac{
P(\Delta S  )}{
P(- \Delta S  )
}
= 
\exp \Delta S 
.
\end{equation}
This also leads to the integral fluctuation theorem~\cite{Shiraishi}, 
$
\langle
\exp ( - \Delta S [ \boldsymbol{\Psi}(0) ; \tau ]  )
\rangle_{\rm ens}
= 
1
$, 
where $\langle \cdots \rangle_{\rm ens}$ represents the ensemble average. 
Moreover, the detailed fluctuation theorem~\cite{Jarzynski} can also be shown to be valid in the present system. 

The ensemble average 
$\langle \Delta S [ \boldsymbol{\Psi}(0) ; \tau ]  \rangle_{\rm ens}$ of the 
stochastic relative entropy 
is never negative, 
which corresponds to the second law of thermodynamics. 
It is given by 
\begin{eqnarray}
& & 
\hspace*{-5mm}
\langle \Delta S [ \boldsymbol{\Psi}(0) ; \tau ]  \rangle_{\rm ens}
=
\int_{-\infty}^{+ \infty} d (\Delta S) \, 
P(\Delta S) \Delta S
\nonumber 
\\
& & 
\hspace*{-5mm}
=
\int 
d \Gamma (\tau) \;
 P[  \boldsymbol{\psi}(\tau)  ; \tau ] 
 \log \left[
 \frac{  P [  \boldsymbol{\psi}(\tau)  ; \tau ] }{
   P  [  \boldsymbol{\psi}(\tau) ; 0 ]
 }
 \right]
 \geq 0
, 
\end{eqnarray}
indicating that $\langle \Delta S [ \boldsymbol{\Psi}(0) ; \tau ]  \rangle_{\rm ens}$ 
is the relative entropy 
(Kullback-Leibler divergence)~\cite{Shiraishi} of the probability distribution 
$P [ \boldsymbol{\psi}(\tau)  ; \tau ]$ at time $\tau$
with respect to 
$P [ \boldsymbol{\psi}(\tau) ; 0 ]$. 
Thus,  $\langle \Delta S [ \boldsymbol{\Psi}(0) ; \tau ]  \rangle_{\rm ens}$ represents 
the information loss incurred when using the initial probability density distribution 
as a surrogate for the true distribution at time $\tau$.


A specific example of the distribution of the initial state vector is given by 
\begin{equation}
\label{initialP}
 P [ \boldsymbol{\psi}(0) ; 0 ]
= 
\frac{1}{Z}
\exp 
\biggl[
- 
\sum_{n=0}^{N_{\rm cvk}  - 1}
\beta_n
| \psi_n (0) |^2
\biggr]
, 
\end{equation}
where 
$
Z 
\equiv 
\int
d \Gamma (0) 
\exp 
\bigl[
- 
\sum_{n=0}^{N_{\rm cvk}  - 1}
\beta_n
| \psi_n (0) |^2
\bigr]
$
and $\beta_n > 0$. 
Note that 
this satisfies 
$
P [ {\bf T} \boldsymbol{\psi}(0); 0 ] 
= 
P [ \boldsymbol{\psi}(0) ; 0 ] 
$ 
and that it becomes stationary,
$
P [ \boldsymbol{\psi}(\tau); 0 ] 
= 
P [ \boldsymbol{\psi}(0) ; 0 ] 
$, 
when all $\beta_n$ takes the same value. 
Here, we assume 
%
$
\beta_n  = \beta_0  / \rho
$
%
for 
$
n = 1, 2, \cdots, N_{\rm cvk} - 1
$, 
where $\beta_0 > 0$ and $0 < \rho  < 1$. 
Then, we obtain 
$
\langle || \boldsymbol{\Psi}(\tau) ||^2 \rangle_{\rm ens}
=
2 \beta_0^{-1}
[ 1 + \rho ( N_{\rm cvk} - 1 ) ]
$ 
and the stochastic relative entropy, 
\begin{equation}
\Delta S [ \boldsymbol{\psi}(0) ; \tau ]
=
\log
\left[
\frac{
P [ \boldsymbol{\psi}(0) ; 0 ] 
}{
P [ \boldsymbol{\psi}(\tau) ; 0 ] 
}
\right]
=
Q
\left( 
\frac{1}{T_{\rm res}} 
-
\frac{1}{T_0}  
\right)
, 
\end{equation}
where 
the decrease in electric field energy per single electron is defined as 
$
Q \equiv (8\pi n_0 L)^{-1}
 \int_{-L/2}^{+L/2} dx  \left(  | E(x, 0) |^2  -  | E(x, t) |^2 \right) 
$
and the effective inverse temperatures of the 
$n=0$ state and other states with $n \geq 1$ 
are given by 
$
1/T_0  \equiv
4 \beta_0  ( 1 + \kappa^2 ) / T
$
and 
$
1/ T_{\rm res} \equiv 1 / (T_0 \, \rho)
$, 
respectively, and 
$
|\psi_0 (\tau) |^2 =  (2 \pi n_0 T L )^{-1} 
( 1 + \kappa^2 )  \int_{-L/2}^{+L/2} dx | E(x, t) |^2 
$
is used. 
Thus, 
$
\Delta S [ \boldsymbol{\psi}(0) ; \tau ]
$
is interpreted as the entropy  
generated per single electron during the time interval $[0, \tau]$
by Landau damping which transfers the electric field 
energy of the $n=0$ state 
with the temperature $T_0$ 
to the thermal reservoirs consisting of the $n\geq 1$ states with the 
lower temperature 
%
$T_{\rm res} = T_0 \, \rho < T_0$. 
%
The fluctuation theorem indicates that either damping or growth of the electric field energy can occur with their relative probabilities constrained by Eq.~(\ref{FT}). 
%
In the nonlinear Vlasov-Poisson system, 
the total energy conservation implies that 
$Q$ equals the increase in kinetic energy 
per single electron.

A total of $10^6$ initial vectors 
$
\boldsymbol{\psi} (0) 
$ 
are randomly generated according to 
$P [ \boldsymbol{\psi}(0) ; 0 ]$
in Eq.~(\ref{initialP}) 
for the numerical verification of the fluctuation theorem. 
Here, 
$\kappa = k \lambda_D = 1/2$, $\rho = 1/20$, and  $N_{\rm cvk} = 20$ are used. 
The normalized mean squared vector components 
$
\beta_0 \langle |\Psi_n (\tau)|^2 \rangle_{\rm ens}
/2
= 
\langle |\Psi_n (\tau)|^2 \rangle_{\rm ens}/
\langle |\Psi_0 (0)|^2 \rangle_{\rm ens}
 $ $(n=0, 1, 2, \cdots, N_{\rm cvk}  - 1)$ 
obtained numerically at $\omega_p t \equiv \tau /(\sqrt{2} \kappa) =  0$, 
0.2, 0.5, 1, 2, and 5 are shown in Fig.~1. 
Figures~2 (a) and (b) show
the probability density function $P(\Delta S)$ of the stochastic relative entropy 
and the ratio $P(\Delta S) / P(-\Delta S)$, respectively,  
at $\omega_p t =  0.2$, 0.5, 1, 2, and 5. 
It is confirmed from Fig.~1 that, for a given time $\tau$, if the value of 
$N_{\rm cvk}$ is taken to be sufficiently large, 
no time evolution is observed in
$\langle | \Psi_n (\tau) |^2 \rangle_{\rm ens}$ 
for large values of $n ( < N_{\rm cvk})$. 
%
For example,   
the results for $N_{\rm cvk}=10$ and $N_{\rm cvk}=20$ are found to be practically identical 
for $\omega_p t \equiv \tau /(\sqrt{2} \kappa) \leq 2$. 
%
Thus, $\langle |\Psi_n (\tau)|^2 \rangle_{\rm ens}$ and 
$P(\Delta S)$ shown for each time in Figs.~1 and 2~(a) can be regarded as equal to
the limiting values to which they converge as $N_{\rm cvk} \rightarrow \infty$. 
%
The fluctuation theorem given in Eq.~(\ref{FT}) is numerically verified in 
Fig.~2~(b) 
%
with 
better accuracy for smaller values of $\Delta S$. 
As $\Delta S$  increases, the value of 
$\exp \Delta S$ grows rapidly, 
requiring a larger number of samples 
to verify the fluctuation theorem with high precision. 
%
Note that 
the fluctuation theorem is valid for any arbitrarily large (but finite) integer $N_{\rm cvk}$, 
and that the actual infinite-dimensional system can be approximated to any desired accuracy 
by a system 
with complex $N_{\rm cvk}$-dimensional state vectors.
Thus, the fluctuation theorem is considered to hold for the infinite-dimensional system 
as the $N_{\rm cvk}\rightarrow \infty$ limit of the finite-dimensional system. 
%

\begin{figure}
\begin{center}
\mbox{}\hspace*{10mm}
\includegraphics[width=60mm]{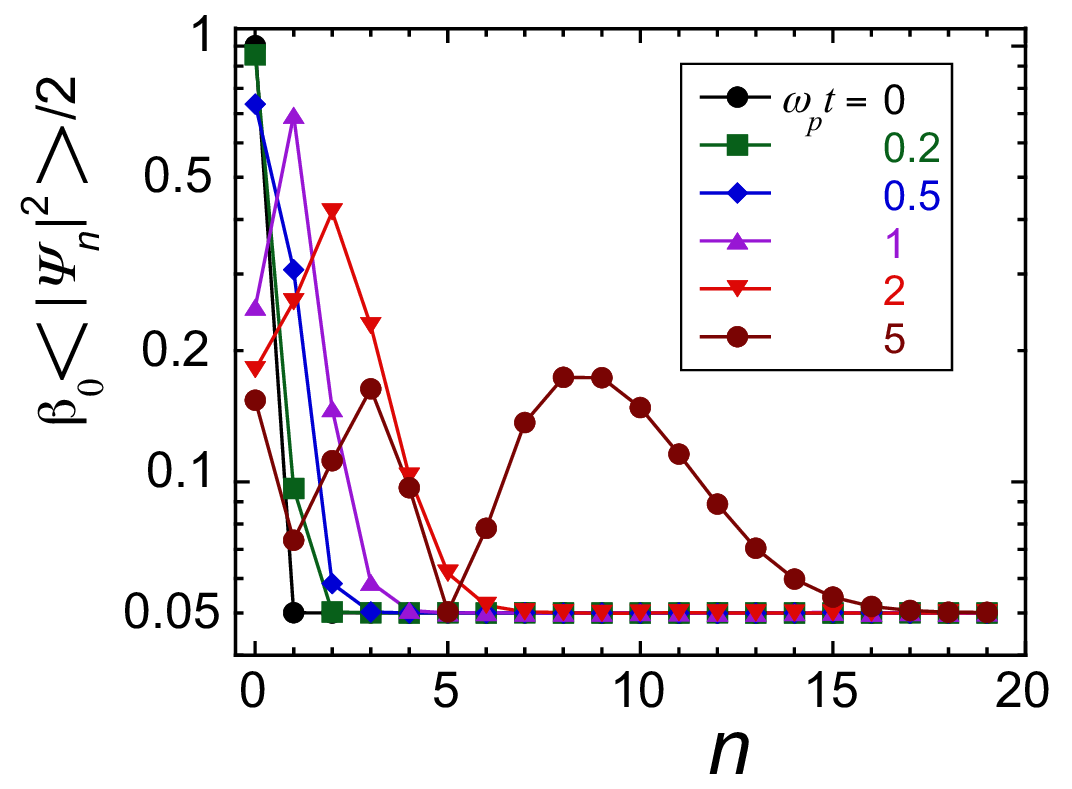}
\\[8mm] \mbox{}
\end{center}
\caption{
 \label{Fig1}
Normalized mean squared vector components 
$\beta_0 \langle |\Psi_n (\tau)|^2 \rangle_{\rm ens}
/2$ 
$(n=0, 1, 2, \cdots, N_{\rm cvk}  - 1)$ 
obtained numerically at time $\omega_p t \equiv \tau /(\sqrt{2} \kappa) =  0$, 0.2, 0.5, 1, 2, and 5. 
 }
\end{figure}

\begin{figure}
\includegraphics[width=50mm]{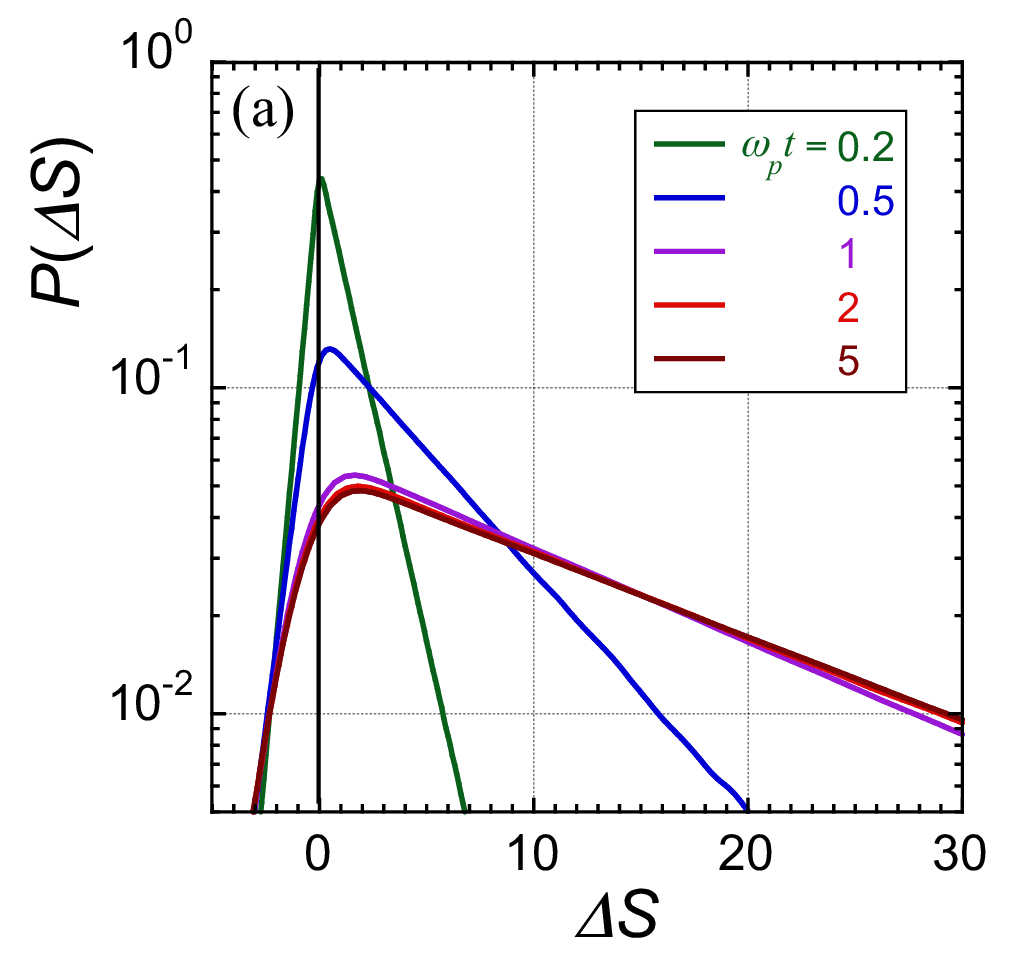}
\\ \mbox{} \hspace*{3mm}
\includegraphics[width=50mm]{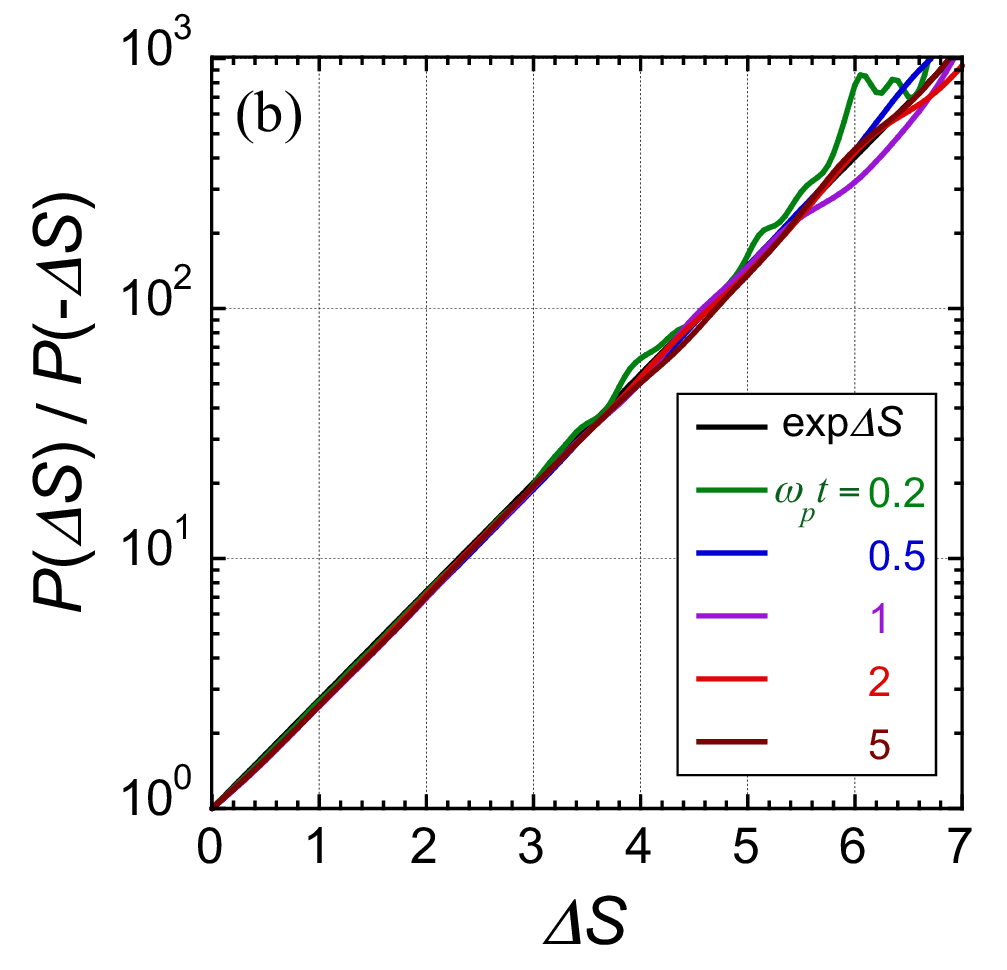}
\\[10mm] \mbox{}
 \caption{
 \label{Fig2}
(a) Probability density function $P(\Delta S)$ of the stochastic relative entropy 
and (b) the ratio $P(\Delta S) / P(-\Delta S)$ 
obtained numerically at time $\omega_p t \equiv \tau /(\sqrt{2} \kappa) =  0.2$, 0.5, 1, 2, and 5.  
}
\end{figure}


This study presents a novel example in which the fluctuation theorem is derived using stochastic relative entropy defined in terms of a probability density functional for a system governed by a kinetic equation with time-reversal symmetry. 
The Schr\"{o}dinger equation and the fluctuation theorem for the Landau damping process presented in this work are expected to contribute to the development of nonequilibrium statistical mechanical formulations of plasma kinetics.  
\\[3mm]
{\bf ACKNOWLEDGMENTS}

The author acknowledges helpful discussions with Prof.\ H.\ Nakamura. 
This work is supported in part by the JSPS Grants-in-Aid for Scientific Research (Grant No.\ 24K07000) 
and in part by the NIFS Collaborative Research Program (No.\ NIFS23KIPT009).
\\[3mm]
%
{\bf AUTHOR DECLARATIONS}
\\
{\bf Conflict of Interest}

The authors have no conflicts of interest to disclose.
\\[3mm]
{\bf Author Contributions}
\\
{\bf Hideo Sugama}: Conceptualization (lead); Data curation (lead); Formal analysis (lead); Funding acquisition (lead); Writing – original draft (lead).
\\[3mm]
{\bf DATA AVAILABILITY}

The data that support the findings of this study are available from
the corresponding author upon reasonable request.



\begin{references}

\bibitem{Landau}
L. D. Landau, 
J. Exp.\ Theor.\ Phys.\ {\bf 16,} 574 (1946).

\bibitem{Case}
K. M. Case, 
Ann.\ Phys. {\bf 7}, 349 (1959).

\bibitem{VK}
N. G. Van Kampen, 
Physica {\bf 21}, 949 (1955).


\bibitem{Nicholson}
D. R. Nicholson, 
{\it Introduction to Plasma Theory} 
(John Wiley \& Sons, New York, 1983), Chap.~6.

\bibitem{H&P}
G. W. Hammett and F. W. Perkins, 
Phys.\ Rev.\ Lett.\ {\bf 64}, 3019 (1990). 

\bibitem{Sugama1999}
H. Sugama,
Phys.\ Plasmas {\bf 6}, 3527 (1999). 

\bibitem{Sugama2006}
H. Sugama and T.-H. Watanabe, 
J. Plasma Phys.\  {\bf 72}, 825 (2006). 

\bibitem{Biancalani}
A. Biancalani, F. Palermo, C. Angioni, A. Bottino, F. Zonca, 
Phys. Plasmas {\bf 23}, 112115 (2016).




\bibitem{Loureiro}
N. F. Loureiro, A. A. Schekochihin, and A. Zocco, 
Phys.\ Rev.\ Lett.\ {\bf 111}, 025002 (2013). 

\bibitem{Plunk}
G. G. Plunk, 
Phys. Plasmas {\bf 20}, 032304 (2013).

\bibitem{Maekaku}
K. Maekaku, H. Sugama and T.-H. Watanabe, 
Phys.\ Plasmas {\bf 31}, 102101 (2024). 

\bibitem{FTES}
D. J.  Evans and  D. J. Searles, 
Adv.\ Phys.\ {\bf 51}, 1529–1585 (2002).



\bibitem{Jarzynski}
C. Jarzynski, 
J.\ Stat.\ Phys.\ {\bf 98}, 77 (2000).

\bibitem{Shiraishi}
N. Shiraishi, 
{\it An Introduction to Stochastic Thermodynamics} 
(Springer Nature, Singapore, 2023), Chap.~5.

\bibitem{Ameri}
A. Ameri, E. Ye, P. Cappellaro, H. Krovi, and N. F. Loureiro, 
Phys.\ Rev.\ A {\bf 107}, 062412 (2023). 

\bibitem{QM}
A. Messiah, 
{\it Quantum Mechanics}, 
(North-Holland, Amsterdam, 1961), Vol.~I.

\end{references}

\end{document}